\documentstyle[amssymb,secnumtab,epsfig]{fbssuppl}

\title{Baryon Resonances in the Double Pion Channel
at Jefferson Lab (CEBAF):
Experimental and Physical Analysis Status and Perspectives }
\author{
M. Ripani\thanks{{\it E-mail address :} ripani@ge.infn.it}
}
\institute{
Istituto Nazionale di Fisica Nucleare, \\
Via Dodecaneso 33, I-16146 Genova (Italy)
}

\begin{document}

\maketitle
\begin{abstract}
The excited baryons made from light quarks are known to
decay in single meson as well as in multimeson final states.
In particular, the double pion production is sensitive to 
many excited states of proton and neutron. Quark models
predict such decays and also that some resonances could
decouple from single meson channels and appear predominantly
in multipion production reactions via electromagnetic
excitation: the so called ``missing resonances''.
These issues are part of the CLAS collaboration scientific program 
at Jefferson Laboratory, where the reaction
$ e N \rightarrow e' N \pi \pi $ is being used in the 
mass region 
between threshold and 2.2 
GeV to investigate baryon resonances and test 
quark models.
In this contribution I will present a framework
for the physical interpretation of the data, especially focusing on
the approach developed by the Genova-Moscow collaboration. 
Some very preliminary raw mass distributions collected with CLAS
are then shown.
\end{abstract}

\section{Introduction}

As established in several years of experimental and theoretical
investigation\cite{Close,Gia90}, mesons and baryons appear to gather 
in mass multiplets
that can be interpreted as the manifestation of ground state
and excitation spectrum
of a system with internal structure. The multiplet structure
is seen as the reflection of symmetry properties of the
Hamiltonian describing the system. Looking at the ground state,
one can see that baryons are organized in an octet of spin 1/2
particles containing proton and neutron, while the well-known 
$\Delta(1232)$ excitation
of the nucleon appears to be member of a spin 3/2 decuplet. 
Octet and decuplet can be in turn put together in a 56-plet,
where the 56 comes from spin states counting. Octet and 
decuplet naturally arise assuming a Hamiltonian symmetric under 
the ``flavour'' group
$SU(3)$ describing the basic $u,d,s$ lightest quarks. 
The spin
$SU(2)$ symmetry does the remaining job, leading to the 56-plet
of $SU(3) \otimes SU(2)$.
Addition of internal quark motion leads to a sequence of orbital 
bands, like those
obtained using an harmonic oscillator confining potential.
Flavour symmetry breaking,
basically due to the mass difference between $u,d$ and $s$ quarks,
leads to the splitting between baryon
states with different strangeness. 
Moreover, to explain the nucleon-$\Delta$ 300 MeV mass 
difference, spin-spin interactions are introduced: they
break SU(2) symmetry, producing a configuration mixing and shifting
the $\Delta$ mass from that of the nucleon, as required.
Finally, the color degree of freedom is assumed to be frozen
in singlet states, such that the resulting hadrons are white,
or colorless.

The use of any quark model incorporating the basic
features of approximate SU(6) symmetry with explicit
flavour-breaking terms and spin-spin interaction, with
a spatial wavefunction obtained from some confining potential, 
is able to account quite reasonably for some general
properties  of baryon states observed experimentally.
In particular the ground state and the first excited states
are usually well accounted for as far as their main static
properties are concerned. However, besides the well-known
discrepancies between electromagnetic properties like
calculated and measured form factors, there is also a major issue
regarding the number of states: the symmetric quark model
predicts a number of states in the second orbital band
which is higher than what seen in experiments. This
is referred to as the problem of ``missing states'' and
stimulated different formulations: in quark models\cite{Kon80}
with hyperfine mixing and explicit meson couplings,
it turns out that some states could have a very weak
pion coupling, while decaying predominatly in multipion
channels, as observed on the other hand in many high-lying
measured states; as the sources of experimental information
are mainly reactions with the pion as projectile or
the single pion as final channel, photoproduced off the
nucleon, it would not be surprising to find that baryon states 
with very small
pion coupling were absent from those data sets. 
Other models\cite{Kon82,Sta93,Cap94,Cap98} based on various meson 
creation assumptions found similar results.
An alternative explanation given for instance by the Quark Cluster
Model \cite{Liu83}
is on the contrary based on a reduction of the
spatial degrees of freedom.
From this introduction, it is quite clear that to test different
model pictures it is necessary to increase the experimental
information on the multipion production, but 
using an electromagnetic probe, to avoid the weak pion
coupling situation that could affect hadron facilities,
without forgetting that the experimental investigation
is made difficult by the often large non-resonant background,
as discussed in the following sections.
Needless to say, Jefferson Lab with the CLAS detector\cite{Dom91,Bur97},
with its high luminosity, acceptance and good momentum
resolution, is the ideal place for performing such kind
of studies: experiments\cite{Nap93} are
currently conducted at Jefferson Lab with namely this goal.

\section{Phenomenology}

\begin{figure}
\epsfig{file=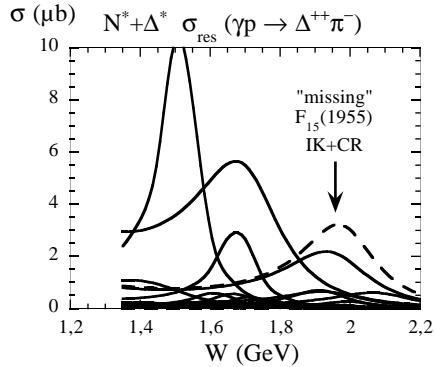,bbllx=0pt,bblly=250pt,bburx=544pt,bbury=550pt,
width=10cm}
\caption{Breit-Wigner cross section in $\gamma p \rightarrow 
\Delta^{++} \pi^{-}$ reaction for known resonances[17] and for
a particular missing state[3,6].  \label{fig:fig1}}
\end{figure}

Main contributions to the double pion production 
are isobar channels like $\Delta(1236) \pi$ and $\rho N$\cite{Cam67}: 
$ e N \rightarrow  e' \Delta \pi \rightarrow  e' N' \pi \pi $, 
$ e N \rightarrow  e' \rho N \rightarrow  e' N' \pi \pi $. All 
isobar production channels can proceed through continuum processes, or 
through the excitation of baryon resonances with a cascade like
$e N  \rightarrow   e' N^{*} \rightarrow   e' \Delta \pi \rightarrow  
e' N' \pi \pi$. 
The double pion production data come mainly from bubble chamber 
experiments with real photons\cite{Cam67,Pia70}, where data about various
charge channels were collected. 
Another experiment at DESY\cite{Eck73} measured 
the electroproduction of $p \pi^{+} \pi^{-}$ off the proton with very 
poor statistics and large binning. Recent photoproduction measurements 
up to slightly above the $D_{13}(1520)$ have been performed at 
Mainz\cite{Bra95,Kru98}, using the DAPHNE large angle detector,
while data in a wider energy range have been collected in Bonn
using the SAPHIR\cite{Kru98,Kle96} detector.
In fig.~\ref{fig:fig1}, data\cite{PDG96} about
known resonance excitations (full curves) together with 
predictions\cite{Kon80,Cap94} for missing states photo-excitation and 
subsequent decay (dashed curve) are used to give an 
estimate of the effect of a missing state: we can expect that 
the cross section should manifest some sensitivity.

\section{Data analysis and interpretation}

At new facilities like CLAS, high luminosity, large geometrical 
acceptance, good efficiency for both charged and neutral particles are 
opening a new era of
unprecedented accuracy in the measurement of exclusive reactions,
allowing a more  
sophisticated data analysis with respect to the past. The main feature 
evident from all the two pion production data collected in 
the past experiments\cite{Cam67,Eck73,Bra95,Kru98} is the presence of 
the isobar "quasi-two-body" states $\Delta \pi$ and $\rho N$. 
A typical approach for separating such 
different isobar contributions is to simply fit their bumps in the 
invariant masses, obtaining approximate cross 
sections. This was the data analysis adopted in most 
of the past experiments with electromagnetic probes\cite{Cam67,Eck73}, 
being interested essentially in the gross 
features and being the data affected by high 
statistical uncertainty. However, the correct 
description of a three-body collision is based on five 
independent kinematical variables in the most general case\cite{Byc73} 
and moreover the isobar quasi-two-body production and 
subsequent decay involves all of them\cite{Pil79}. 
	Investigations of double pion production from pion 
beams have been 
in fact conducted using isobar model approaches containing the partial 
wave 
expansion for each quasi-two-body process and fitting the data in the 
full kinematical space\cite{Man84}. 
Any resonance analysis with the goal of extracting the baryon resonance decay 
branches in a quasi-two-body channel or the product of the e.m. transition
matrix elements with the strong decay one (the ``electrostrong properties''
\cite{Muk98}), in a way as model independent as possible,
needs such an isobar partial wave separation from the data as 
an input, similar to what done in previous analysis\cite{Ber75,Arn96}.  
Therefore in a preliminary simple study done on the 
$\Delta \pi$  channel, pseudo-events 
were generated using only the geometrical partial wave expansion\cite{Jac59},
with no explicit dynamics, then refitted to retrieve the partial wave 
coefficients. 
The outcome was that even in this simple case the fitting code was not 
able to retrieve the large number of independent helicity 
amplitudes that arise with increasing angular momentum. 
Different solutions could be in principle pursued: one way is 
to add polarisation observables, 
in order to have a more constrained fit; a second possibility is to 
use orbital 
waves constrained by threshold behavior; a further possibility
is to use simple model assumptions
for the continuum and the resonances. In fact, it is important
to consider that the \(N^{*}\) study in two-pion production is 
affected by strong non-resonant processes and therefore model-independent 
methods of analysis may be not effective. For all these reasons
the choice in the Genova-Moscow collaboration\cite{Mok98} was to 
give up the requirement of minimal model dependence and
use some partial wave content suggested from a model as input
to the analysis, as described in the next section.

\section{Our approach for the quasi-two-body channels}

After the old work that followed the bubble chamber first experiments\cite{Cam67,Bar72},
recent approaches to describe double pion photoproduction have
been presented in a few papers\cite{Gom94,Mur96} based
on a variety of tree-level diagrams and a few baryon resonances.
The restricted number of resonances included however makes 
them strictly applicable only for W lower than 1.6 - 1.7 GeV; 
moreover non-resonant 
terms have been evaluated only at the photon point and not always
corrected for unitarity absorption effects.

The Genova-Moscow approach to calculate cross sections is described in more detail
in \cite{Ang98,Mok98}. I report here the general features.
Following the data, we also use a coherent superposition of 
\(\gamma_{r,v} p\rightarrow \pi^{-}\Delta^{++}\)\(^{1}\) and 
\(\gamma_{r,v} p\rightarrow \rho p\)\(^{1}\) quasi-two-body subchannels.
All remaining processes are described in phase space approximation.
\footnotetext[1]{Indexes r,v stand for real and virtual photons respectively.}
The \(\gamma_{r,v} p\rightarrow \pi^{-}\Delta^{++}\) reaction is described by
a superposition of \(N^{*}\), \(\Delta^{*}\) excitation in s-channel and a
minimal set of non-resonant processes obeying gauge invariance conditions,
similar to what done in the previous literature\cite{Cam67,Bar72}. 
Non-resonant amplitudes are derived from an
effective Lagrangian\cite{Bar72}, as done for other meson production 
channels\cite{Ben95}. 
New features of this approach are: (1) the treatment 
of particle's off-shell behaviour through introduction of vertex
functions that result from a combination of electromagnetic
form factors and strong form factors specified via a cut-off
parameter\cite{Ang98,Mok98}; data\cite{Beb78}
have been used to determine part of them, 
while the remaining terms in the calculation were
derived imposing gauge invariance\cite{Ang98,Mok98};
(2) the initial and final state absorption due to
competitive channels follows \cite{Got64}, but the elastic hadronic amplitudes
are reconstructed using resonant
contributions taken from \cite{Man92}, plus a smooth
background parametrised in the same fashion of \cite{Dyt97}.
It is important to stress here that ``missing'' resonances
with strong two pion coupling
should be introduced consistently in both the e.m.
amplitudes as well as in the absorption:
this evaluation is currently under way in our Genova-Moscow
collaboration. 
\begin{figure}
\epsfig{file=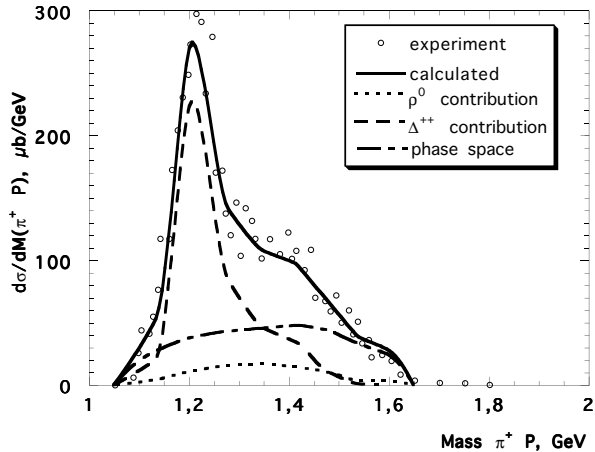,bbllx=0pt,bblly=200pt,bburx=544pt,bbury=550pt,
width=10cm}
\caption{Invariant mass distribution for $p \pi^{+}$ pair from SAPHIR 
data[36] and the Genova-Moscow fit: the meaning of the different contributions
is reported in the picture legend .  \label{fig:fig3}}
\end{figure}
Results from this calculation for
\(\gamma_{r,v} p\rightarrow \pi^{-}\Delta^{++}\) reaction
are extensively reported in the contribution presented by
V. Mokeev at this Workshop. Basically, main findings are that
leaving the absorption
as a free parameter it is possible to get a very good fit of
the data, but resonance extraction becomes more uncertain;
using the above parametrisation
of initial and final state interaction,
data are not completetely reproduced\cite{Mok98},
but this discrepancy opens room to interesting effects like
possible missing states contributions.

As the experiment does not measure isobar production directly of course,
but only the two pion final state, in order to have a complete tool
for the analysis next step was to merge together the $\Delta \pi$ and $\rho N$
production channels plus a phase space in a full three-body
calculation. A new feature of the Genova-Moscow approach in this respect is
the introduction of decay strong form factors for the $\Delta$
and $\rho$ decay\cite{Cas81}. 
 In fig.~\ref{fig:fig3}, I present an example of the results 
obtained fitting invariant
mass distributions from recent photoproduction data\cite{Kle96}. 
The fit is pretty good, therefore providing a
promising tool for a quite reliable extraction
of the different isobar components in the reaction.
%
%

\section{A quick look at the first CLAS data}

\begin{figure}
\epsfig{file=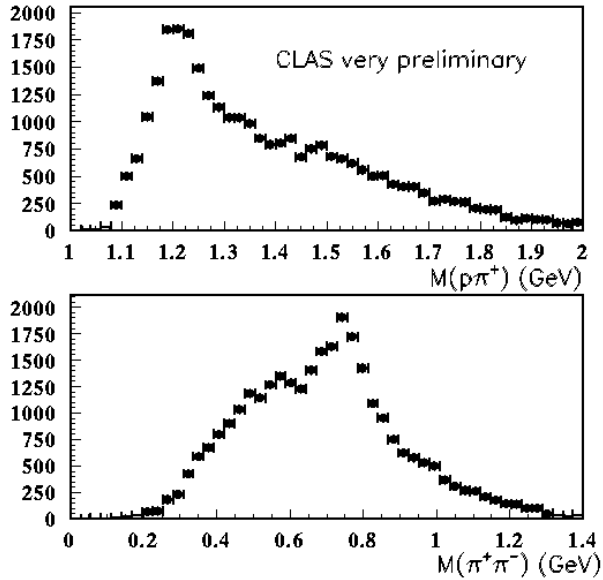,bbllx=0pt,bblly=200pt,bburx=524pt,bbury=590pt,
width=10cm}
\caption{Invariant mass distributions for the $p \pi^{+}$ (top) and the
$ \pi^{+}  \pi^{-}$ (bottom) pairs from CLAS for W>1.7 GeV. These are raw data
without any energy or momentum transfer binning; acceptance 
correction were also not applied.   \label{fig:fig4_mod}}
\end{figure}
In this talk I showed also some very preliminary data from CEBAF-CLAS experiment
E-93-006. In fig.~\ref{fig:fig4_mod} a snapshot from 
a sample of CLAS data is reported,
showing the invariant mass distributions for the $p \pi^{+}$ and
the $ \pi^{+}  \pi^{-}$ pairs for W>1.7 GeV. The data were neither binned in W
nor in $Q^{2}$ and they were not corrected for the detector acceptance;
therefore they represent only the raw output from CLAS with the
intent of giving essentially an idea of its capabilities. However, 
the contributions of the $ \Delta^{++} $ and of the
$ \rho^{0} $ meson, respectively, are recognizable. The data already collected contain
about an order of magnitude more events and nearly the same amount
will be accumulated in other planned running periods, therefore allowing 
a quite large binning and investigation of details such as decay angular
distributions with much higher accuracy than the past.
%
%

\section{Summary and conclusions}

New experiments like those currently conducted at Jefferson Laboratory
are providing a wealth of new accurate data about exclusive electromagnetic
reactions. Two pion production is one of the main subjects of
investigation, being related to baryon resonances coupled to this channel.
A specific approach for the isobar channels that appear
in the two pion production has been developed in the framework of
the Genova-Moscow collaboration, taking particular care about the $\Delta \pi$,
channel description, especially concerning initial and final state
absorption, gauge invariance,
vertex functions and multiple resonances. $\rho$ meson production was
instead described through a simple diffraction ansatz. These calculations 
are able to
give good account of existing data about differential cross sections
and invariant mass distributions, therefore promising to allow a
quite complete data analysis and a first evaluation of resonance
contributions. In the first data from CLAS it is already possible to
recognize the isobar formation with good statistics,
opening the route to more detailed studies of the involved dynamics.



\makeatletter \if@amssymbols%
\else\relax\fi\makeatother

\SaveFinalPage
\end{document}